\documentclass[prb,groupedaddress,twocolumn,showpacs]{revtex4-2}

\usepackage{graphicx}
\usepackage{dcolumn}
\usepackage{bm}
\usepackage{hyperref}
\usepackage{url}
\usepackage{color}
\usepackage{amsmath}
\usepackage[running]{lineno}

\newcommand{\pcto}{PbCuTe$_{2}$O$_{6}$\,}
\newcommand{\scto}{SrCuTe$_{2}$O$_{6}$\,}

\newcommand{\jin}{$J_{intra}$\,}

\begin{document}
\title{Weak three-dimensional coupling of Heisenberg quantum spin chains in \scto}

\author{S. Chillal}
\email[*]{shravani.chillal@helmholtz-berlin.de}
\affiliation{Helmholtz-Zentrum Berlin f\"ur Materialien und Energie, Hahn-Meitner Platz 1, 14109 Berlin, Germany}

\author{A. T. M. N. Islam}
\affiliation{Helmholtz-Zentrum Berlin f\"ur Materialien und Energie, Hahn-Meitner Platz 1, 14109 Berlin, Germany}

\author{P. Steffens}
\affiliation{Institut Laue-Langevin, 71 Avenue des Martyrs, 38042 Grenoble Cedex 9, France}

\author{R. Bewley}
\affiliation{ISIS Facility, STFC Rutherford Appleton Laboratory, Oxfordshire OX11 0QX, UK}

\author{B. Lake}
\affiliation{Helmholtz-Zentrum Berlin f\"ur Materialien und Energie, Hahn-Meitner Platz 1, 14109 Berlin, Germany}
\affiliation{Institut f\"ur Festk\"orperphysik, Technische Universit\"at Berlin, Hardenbergstr. 36, 10623 Berlin, Germany}

\begin{abstract}
The magnetic Hamiltonian of the Heisenberg quantum antiferromagnet \scto is studied by inelastic neutron scattering technique on powder and single crystalline samples above and below the magnetic transition temperatures at 8~K and 2~K. The high temperature spectra reveal a characteristic diffuse scattering corresponding to a multi-spinon continuum confirming the dominant quantum spin-chain behavior due to the third neighbour interaction $J_{intra}=4.22$~meV ($49$~K). The low temperature spectra exhibits sharper excitations at energies below 1.25~meV which can be explained by considering a combination of weak antiferromagnetic first nearest neighbour interchain coupling $J_{1}=0.17$~meV ($1.9$~K) and even weaker ferromagnetic second nearest neighbour $J_{2}=-0.037$~meV ($-0.4$~K) or a weak ferromagnetic $J_{2}=-0.11$~meV ($-1.3$~K) and antiferromagnetic  $J_{6}=0.16$~meV ($1.85$~K) giving rise to the long-range magnetic order and spin-wave excitations at low energies. These results suggest that \scto is a highly one-dimensional Heisenberg system with three mutually perpendicular spin-chains coupled by a weak ferromagnetic $J_2$ in addition to the antiferromagnetic $J_1$ or $J_6$ presenting a contrasting scenario from the highly frustrated hyper-hyperkagome lattice (equally strong antiferromagnetic $J_1$ and $J_2$) found in the iso-structural \pcto. 
\end{abstract}


\date{\today}

\maketitle

Low-dimensional Heisenberg magnetic systems are home to a rich variety of exotic magnetic properties. Especially so when the constituents are made of the smallest spin unit ($S=\frac{1}{2}$), ranging from the disordered ground state in a one dimensional (1D) chain of spins giving rise to peculiar excitations with a quantum number of $S=\frac{1}{2}$ known as spinons~\cite{Lake2005,Mourigal2013,Lake2013} to the highly entangled quantum spin liquid ground state~\cite{Han2012,Balz2016} when the spins are arranged in frustrated motifs such as the two-dimensional (2D) network of triangles known as the {\it Kagome} lattice. However, these ideal conditions are rarely present in real systems where additional magnetic interactions such as an unfrustrated or frustrated inter-chain (or inter-layer) coupling~\cite{Hase1993,Schapers2013}, anisotropy~\cite{Kataev2001} or a Dzyaloshinskii-Moriya (DM) interaction~\cite{Coldea2002} give rise to a variety of long-range ordered (LRO) magnetic structures at lower temperatures. Consequently, the lower energy part of the spinon excitation spectra is also modified giving insights into the fundamental effect of various magnetic interactions.

\begin{figure}
\includegraphics[width=1 \columnwidth]{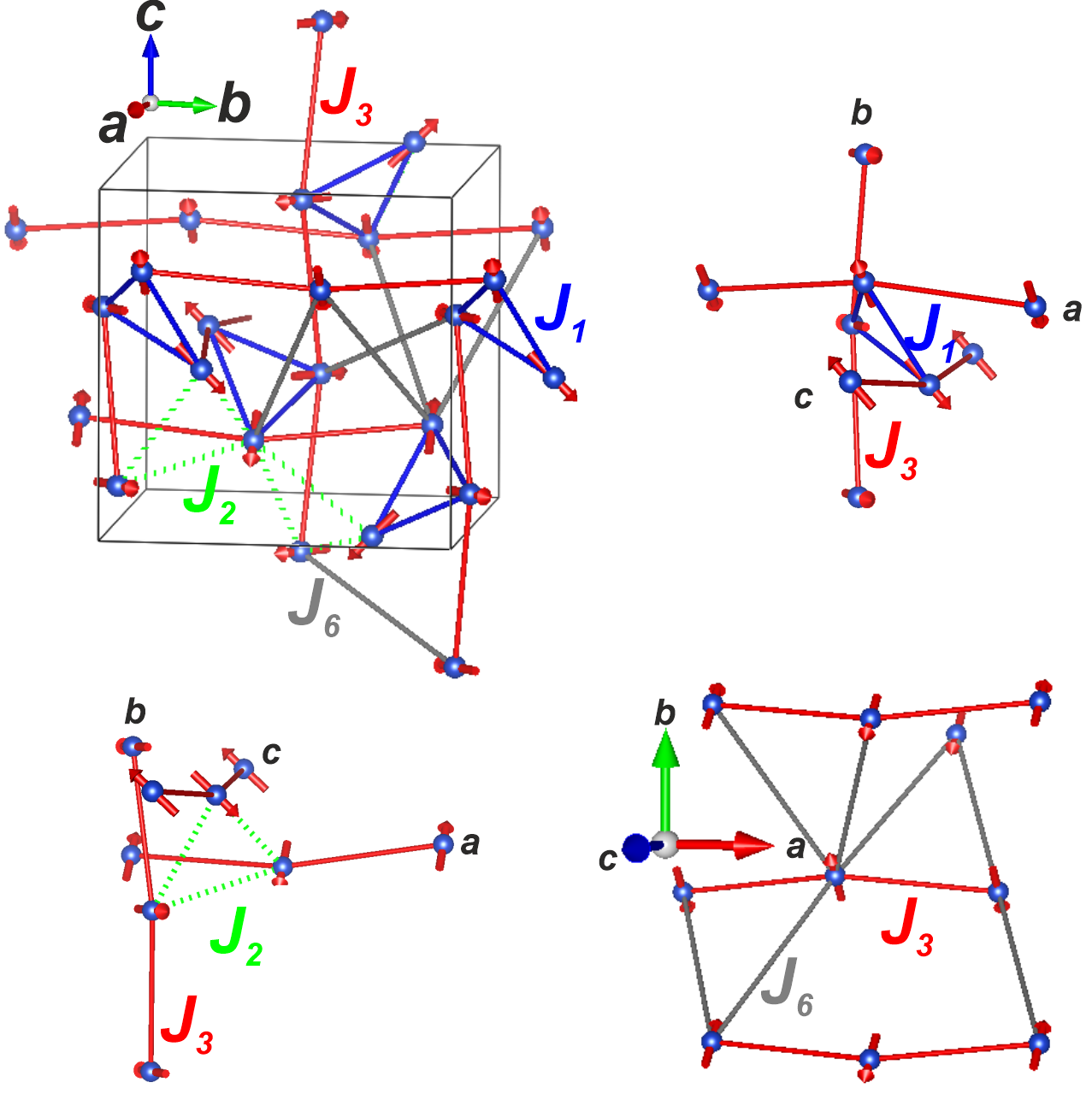}
\caption{Magnetic structure of \scto showing only the Cu$^{2+}$ ions and their spin directions. The isolated triangles formed by the $J_{1}$ interaction (blue lines) and the hyperkagome lattice formed by the $J_{2}$ interaction (green lines) couple the three mutually perpendicular chains along the cubic  {\it a}, {\it b} and {\it c} axes (red lines). Grey lines indicate an additional 6th nearest neighbor interaction $J_{6}$ proposed by DFT~\cite{Ahmed2015,Bag2021}.}
\label{Fig1}
\end{figure}

ACuTe$_{2}$O$_{6}$ (A = Pb, Sr, Ba) are a family of compounds crystallizing in the cubic structure (space group {\it P4$_{1}$32}) and have a single Wyckoff site ($12d$ multiplicity) for the magnetic Cu$^{2+}$ ion. As depicted in the schematic Fig.~\ref{Fig1}(a), the 1st, 2nd and 3rd nearest neighbour interactions between the Copper ions lead to isolated triangles ($J_{1}$), corner-shared triangles known as the hyperkagome lattice ($J_{2}$) and one dimensional chains ($J_{3}$) respectively. In the presence of antiferromagnetic interactions between the spins, the $J_{1}$ and $J_{2}$ give rise to geometrical frustration in the system whereas the $J_{3}$ forms three mutually perpendicular antiferromagnetic chains parallel to the cubic {\it a}, {\it b} and {\it c} axes respectively. Therefore, depending on the ratio of $J_{1}$ and $J_{2}$ these systems can host many interesting magnetic properties in their ground state. The quantum spin liquid state found in \pcto is especially important~\cite{Koteswararao2014,Khuntia2016,Chillal2020}. Electronic band structure calculations using density functional theory (DFT) suggest that the equally strong and highly frustrated $J_{1}$ and $J_{2}$ are responsible for this rare ground state hosted on a three-dimensional network of corner-shared triangles known as the hyper-hyperkagome lattice~\cite{Chillal2020}. Whereas, the role of unfrustrated chain interaction $J_3$ with half the strength of the frustrated $J_{1}$ or $J_{2}$ is thought to drive the system closer to a magnetically ordered state. The ratio of these interactions ($J_{1}:J_{2}:J_{3}:J_{4}\sim1:1:0.5:0.2$, where $J_{4}$ is an additional chain interaction) was successfully verified by reproducing the experimental dynamic neutron structure factor~\cite{Chillal2020}. However, we note that the absolute values of the interactions have not been confirmed experimentally. Similarly, there have been multiple reports on the values of interactions using DFT for the magnetically ordered Sr and Ba variants of this family~\cite{Ahmed2015,Bag2021}. Although the strongest interaction predicted by DFT has been measured experimentally, the magnitude and nature of the weaker interchain couplings responsible for the magnetic order have not been confirmed.  In this context, experimental investigation of magnetic excitation spectra of these compounds is essential to understand the interplay of the intra-chain and the frustrated interchain interactions in determining the magnetic phase diagram. On the other hand, it will also enable an independent verification of the applicability of the DFT method for these quantum spin systems. Hence, here we investigate the magnetic excitations of \scto using inelastic neutron scattering.

\scto has a lattice constant of {\it a}=12.4373~\AA~\cite{Wulff1997}. The DC susceptibility of single crystal and polycrystalline \scto yields a negative Curie-Weiss temperature of $\theta_{\rm CW}=-27\pm1$~{\rm{K}}~\cite{Chillal2020b} revealing predominantly antiferromagnetic exchange interactions~\cite{Koteswararao2015,Ahmed2015,Chillal2020b}, and shows a broad maximum at 32~K. This feature has been attributed to a one-dimensional spin-$\frac{1}{2}$ Heisenberg antiferromagnetic chain revealing $J_{3}=49$~K~\cite{Koteswararao2015,Ahmed2015,Chillal2020b} as the dominant interaction. However, two sharp features occur in the susceptibility at lower temperatures T$_{N1}=5.5$~K and T$_{N2}=4.5$~K, where a sharp $\lambda$-type anomaly is also observed in the heat capacity, indicating the onset of long-range magnetic order in the system. These anomalies reveal non-negligible frustrated inter-chain coupling due to the finite $J_1$ and $J_2$~\cite{Koteswararao2015,Ahmed2015}. In addition, the compound exhibits magneto-dielectric coupling at T$_{N1}$ and T$_{N2}$~\cite{Koteswararao2016} attributed to the non-centro-symmetric nature of the structural symmetry. Furthermore, specific heat, magnetization and dielectric constant measurements as a function of applied magnetic field reveal a complex phase diagram with additional field-induced phases~\cite{Koteswararao2015,Ahmed2015}. 

In the low-temperature magnetic phase, the spins order in a 120$^{\circ}$ co-planar structure around the triangles formed by $J_1$~\cite{Saeaun2020,Chillal2020b} with magnetic propagation vector $q=(0,0,0)$. The Cu$^{2+}$ spins on the three vertices of these triangles are aligned along the local $(110)$ directions and couple together the three mutually perpendicular chains formed by $J_3$ along the cubic {\it a}, {\it b} and {\it c} axes. This result suggests that $J_1$ is antiferromagnetic and responsible for the interchain coupling. On the other hand, the earlier density functional (DFT) band-structure calculations suggest that the contribution of $J_1$ ($\sim0.025$~meV) to the interchain coupling is negligible whereas the frustrated antiferromagnetic second and sixth nearest neighbours $J_2$ ($\sim 0.34$~meV) and $J_6$ ($\sim0.17$~meV, represented in fig.~\ref{Fig1}), respectively, are responsible for the magnetic ordering~\cite{Ahmed2015}. In a recent calculation~\cite{Bag2021} it is also found that the $J_1$ is absent while interchain coupling is led by ferromagnetic $J_2$ ($\sim -0.4$~meV) and antiferromagnetic sixth nearest neighbor $J_6$ ($\sim0.17$~meV). Although the bulk magnetic and thermodynamic properties of \scto give an estimation of this inter-chain coupling, the experimental confirmation of the magnetic Hamiltonian including the true nature of the interchain coupling is still unclear. 

In this paper, we present a detailed investigation into the magnetic Hamiltonian of \scto using inelastic neutron scattering of polycrystalline and single crystal samples. Our results indicate a spinon continuum resembling that of the excitations observed in the prototypical 1D spin chain systems. The intra-chain interaction $J_{intra}$ calculated from the lower bound of the continuum is 4.22~meV, in agreement with the magnetic susceptibility. Below the magnetic transition temperature, spin waves appear at an energy transfer lower than 1.25~meV which can be described by two independent inter-chain models: first combination consists of a weak antiferromagnetic first neighbor interaction $J_1=0.17\pm0.025$~meV and weaker ferromagnetic second nearest neighbor interaction $J_2=-0.037\pm0.012$~meV and second, is described a ferromagnetic $J_2=-0.11\pm0.002$~meV and ferromagnetic $J_6=0.16\pm0.004$~meV. 

\section{Samples \& Experimental Methods}

The inelastic neutron scattering of \scto was investigated on the polycrystalline and single crystal samples that were prepared following the procedure explained in Ref.~\cite{Chillal2020b}.

 The powder inelastic neutron scattering data was obtained at the time-of-flight spectrometer LET located at the ISIS facility, Didcot, United Kingdom. For these measurements the polycrystalline powder was filled (weight $10$~g) in an Aluminium can. The measurements were performed at $T=8$~{\rm K} and $T=2$~{\rm K} with incident energies: $E_i=18.1$, $5.64$, $2.72$, $1.59$~meV. Single crystal inelastic neutron measurements in the $\lbrack{h,h,l}\rbrack-$plane were obtained at the ThALES triple-axis spectrometer~\cite{ILLdata} using the flatcone detector at the ILL, Grenoble, France. The wavevector maps at constant energy were measured on ThALES at $T=2$~{\rm K} while rotating the crystal in $0.2~\deg$ steps with a fixed final energy of $E_i=4.06$~meV giving an energy resolution $0.123$~meV. The wavevector resolution in the plots is $0.05$~r.l.u${\times}0.05$~r.l.u. 
 The full scattering cross-section $S(\mathbf{Q},\omega)$ was measured at the LET spectrometer~\cite{ISISdata} in ISIS at $T=2$~{\rm K} and $8$~{\rm K} with incident energies of $E_i=19.1$~meV, $9.63$~meV, $5.8$~meV, $3.87$~meV and $2.76$~meV. At each of the incident energies, the energy resolution is approximately 3\% of the $E_i$. The experimental data has been reduced and analysed using the Matlab-based packages HORACE~\cite{Ewings2016} and SPINW~\cite{Toth2015} respectively.
 
\section{Results}

\subsection{Powder inelastic neutron scattering}

\begin{figure}
\includegraphics[width=1 \columnwidth]{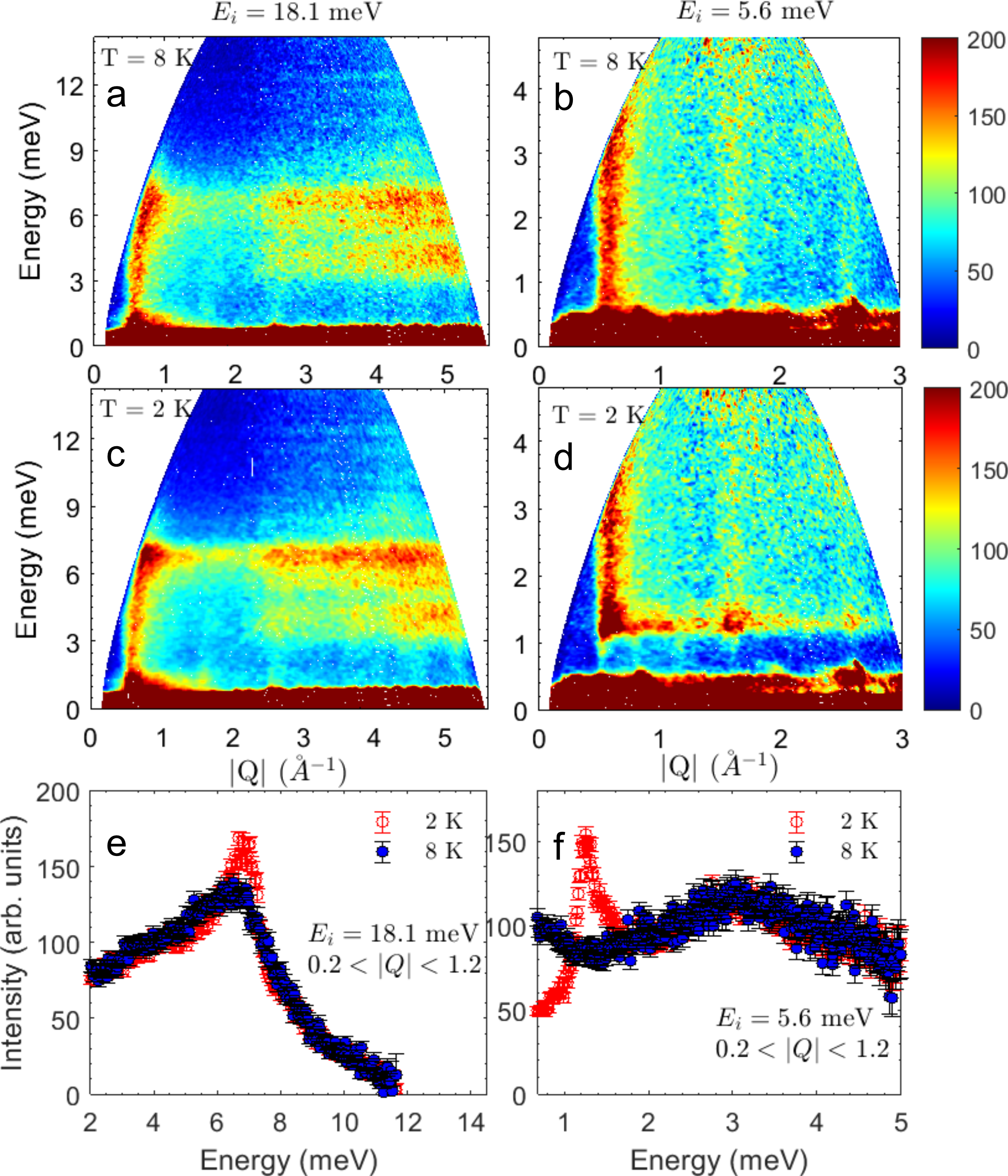}
\caption{\noindent Excitation spectra obtained on the powder of \scto using the time-of-flight spectrometer LET with an incident energy of $E_{i}=18.1$~meV and $E_{i}=5.6$~meV respectively at: \textbf{(a)}-\textbf{(b)} $T=8$~{\rm K} and \textbf{(c)}-\textbf{(d)} $T=2$~{\rm K}. These plots clearly show excitations up to 9~meV at 8~K and 2~K as well as the modified low-energy part of the excitations below the ordering temperature. \textbf{(e)}-\textbf{(f)} shows the integrated intensity of the magnetic excitations plotted as a function of energy for $E_{i}=18.1$~meV and $E_{i}=5.6$~meV (resolution $\delta{E}=1.3$~meV and  0.23~meV) integrated over the wavevector range $0.2\leq|\mathbf{Q}|\leq1.2$ \AA$^{-1}$).}
  \label{Fig2}
  \end{figure}

Figure~\ref{Fig2}(a)-(d) show inelastic neutron scattering (INS) spectra of the powder sample of \scto measured on the time-of-flight spectrometer LET with two incident energies ($E_i=18.1, 5.6$~meV with energy resolution $\delta E=1.26, 0.23$~meV respectively) above and below the ordering temperatures at 8~K and 2~K respectively. The inelastic spectrum at 8~K shows a streak of diffuse scattering originating at the wave vectors $|\mathbf{Q}|=0.502, 1.504, 2.504~\AA^{-1}$ [in Fig.~\ref{Fig2}(a)-(b)] which correspond to the wavevectors of the antiferromagnetic Bragg peaks $(100), (300), (500)$ respectively observed in the neutron diffraction suggesting the dispersive nature of the excitations. As shown in Fig.~\ref{Fig2}(a), this inelastic intensity is clearly visible up to $\sim9$~meV and forms a flat band. Furthermore, the intensity strongly decreases with increasing wave vector transfer, typical of the magnetic form-factor. The energy dependence of this magnetic excitation intensity, plotted in Fig.~\ref{Fig2}(e) (blue circles), peaks up at 6.7~meV and extends up to 12~meV. This is reminiscent of the spinon-continuum with lower bound occurring at $\frac{\pi}{2}$\jin and upper bound at $\pi$\jin which results in a \jin$=4.22$~meV in \scto considering the peak as the lower bound of the continuum. Additional flat bands are also observed in the spectrum in Fig.~\ref{Fig2}(a) at higher wave-vector transfer $|\mathbf{Q}|\geq2.504~\AA^{-1}$ whose intensity increases with $|\mathbf{Q}|$ suggesting their origin to be phononic in nature. 

In the magnetically ordered state at 2~K, the spinon-like diffuse streaks at the $|\mathbf{Q}|=0.502, 1.504, 2.504~\AA^{-1}$ (in Fig.~\ref{Fig2}(c)) remain unchanged. However, we observe that the intensity at $\frac{\pi}{2}$\jin$=6.73$~meV is stronger as well as sharper than at $8$~K [see red circles in Fig.~\ref{Fig2}(e)] suggesting a modification of the structure factor of the spinon continuum due to the magnetic order. Furthermore, the spectrum with finer energy resolution, presented in Fig.~\ref{Fig2}(d), reveals additional modifications in the low-energy part in the form of a flat band at $\sim1.25$~meV and sharp streaks below it at $|\mathbf{Q}|=0.502, 0.816, 1.16, 1.5~\AA^{-1}$. However, no difference is observed above the flat band (see Fig.~\ref{Fig2}(f)) with respect to the intensity at higher temperature.

\subsection{Single crystal inelastic neutron scattering} \label{secB}


\begin{figure}
\includegraphics[width=1 \columnwidth]{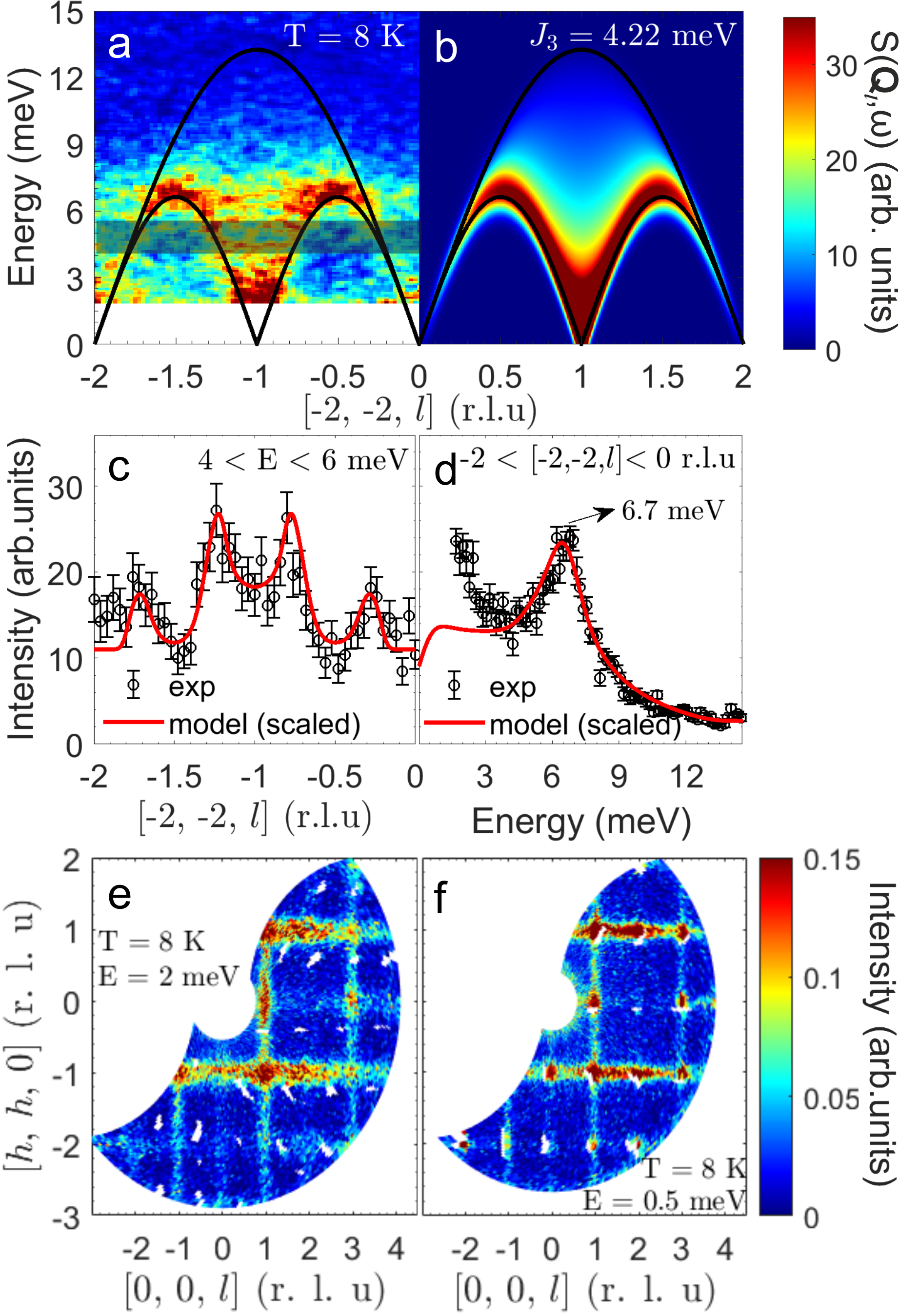}
\caption{\textbf{(a)} Excitation spectra obtained on the single crystal of \scto using the time-of-flight spectrometer LET, with an incident energy of $E_{i}=19.1$~meV. The intensity includes $\pm0.5$~r.l.u in the vertical scattering plane ($\lbrack{h,-h,0}\rbrack$) as well orthogonal in-plane scattering ($\lbrack{h,h,0}\rbrack$). \textbf{(b)} the simulated dynamical structure factor $S_{1D}(\mathbf{Q}_{l},\omega)$ of a spin-1/2 Heisenberg antiferromagnetic chain considering $J_3=$\jin$=4.22$~meV. \textbf{(c)} The distribution of magnetic intensity across the Brillouin zone integrated over an energy transfer of $4<E<6$~meV. The non-zero intensity at $-1.5\leq[-2,-2,l]\leq-0.5$~r.l.u confirms the continuum nature of the excitations. \textbf{(d)} shows that the magnetic excitations survive up to $\approx13$~meV. \textbf{(e)-(f)} Single crystal spectra measured on the ThALES spectrometer in the $\lbrack{h,h,l}\rbrack-$plane at constant energy transfers of E$=2$~meV, $0.5$~meV respectively. Non-magnetic features such as Bragg peak tails have been removed from the spectrum.}
\label{Fig3}
\end{figure}

In order to better understand the magnetic excitations of \scto, detailed inelastic neutron scattering data were collected on a single crystal sample at the ThALES flat-cone spectrometer as well as the LET time-of-flight spectrometer. Figure ~\ref{Fig3}(a) presents the excitation spectrum as a function of $[-2,-2,l]$ along the cubic {\it c}-axis collected for an incident energy of $E_{i}=19.1$~meV at temperature $T=8$~K. The spectrum reveals a dispersive, diffuse pattern with maximum intensity originating at the antiferromagnetic zone-center ($l=-1$) which clearly extends up to $9$~meV. Figure~\ref{Fig3}(c) shows a cut along the $l-$direction integrated over an energy range $4\leq E\leq6$~meV indicating a non-zero intensity around the zone-center at $l=-1$. Additional clearly separated shoulder peaks at $l=-1.75,-0.25$ are also visible on either side of the zone-center. These observations further confirm the diffuse nature of the excitations. 
Figure~\ref{Fig3}(d), where intensity in the whole Brillouin zone is summed and plotted as a function of energy, shows that the excitations peak up at the energy transfer of $\sim6.7$~meV and survive up to $\sim13$~meV. These observations are consistent with the powder sample and suggest that the excitations are due to the presence of 1D spin-1/2 chains in the compound which result in fractional excitations known as spinons. 

Considering the chain interaction \jin$=4.22$~meV, the dynamic structure factor (DSF) of a 1D Heisenberg spin$-\frac{1}{2}$ chain, $S_{1D}(\mathbf{Q}_{l},\omega)$ is calculated for the Hamiltonian:
\begin{equation}
\mathcal{H}_{intra}=\sum_{k=(i<j)}\j_{intra} \mathbf{S}_{i}\cdot\mathbf{S}_{j},
\label{eq1}
\end{equation}
using the algebraic Bethe-ansatz method~\cite{Caux2005,Caux2006}. 
Figure~\ref{Fig3}(b) shows the $S_{1D}(\mathbf{Q}_{l},\omega)$ convoluted with a Gaussian of full width half maximum, FWHM$=1.26$~meV to account for the instrumental energy resolution. This structure factor reproduces the main features of the excitation spectrum such as the lower bound of the 2-spinon continuum (sinusoidal dispersion with amplitude $E_{lower}=\frac{\pi}{2}$\jin) and the gradually decreasing intensity as a function of energy which vanishes at the upper boundary $E_{upper}=\pi$\jin. These boundaries are clearly indicated by the black solid lines in Fig.~\ref{Fig3}(a)-(b). To compare the theoretical intensities with the energy and wave vector cuts through the data in Fig.~\ref{Fig3}(c)-(d)), the same cuts are taken through the theory (red solid lines).

The constant energy plots, shown in the Fig.~\ref{Fig3}(e)-(f) for the $[h,h,l]$ scattering plane at $2$~meV and $0.5$~meV reveal mutually perpendicular streaks of intensity along $[h,h,0]$ and $[0,0,l]$ directions. In a single Heisenberg antiferromagnetic spin-1/2 chain system, the intensity modulates with the sinusoidal dispersion along the direction of chain $[0,0,l]$ (as shown in Fig.~\ref{Fig3}(a)-(b) for \scto), but it is non-dispersive perpendicular to the chain due to the absence of long-range correlations in the other two directions giving rise to uniform streaks along $[h,h,0]$ at odd values of $l$. However, strong rods of intensity are also observed perpendicular to the $[h,h,0]$ direction, as shown in Fig.~\ref{Fig3}(e)-(f), suggesting the contribution of streaks from the chains parallel to the $h,k-$directions resulting in strongest intensity at the $[h,h,h]$ nodes where the intensity from all the three chains are superposed. Therefore, this data confirms the presence of three mutually perpendicular spin-1/2 chains in \scto due to the third nearest neighbour interaction $J_3$ running parallel to the cubic axes as shown by the red lines in Fig.~\ref{Fig1}(a).   

\subsection{Magnetic Hamiltonian}

\begin{figure*}
\includegraphics[width=2.08 \columnwidth]{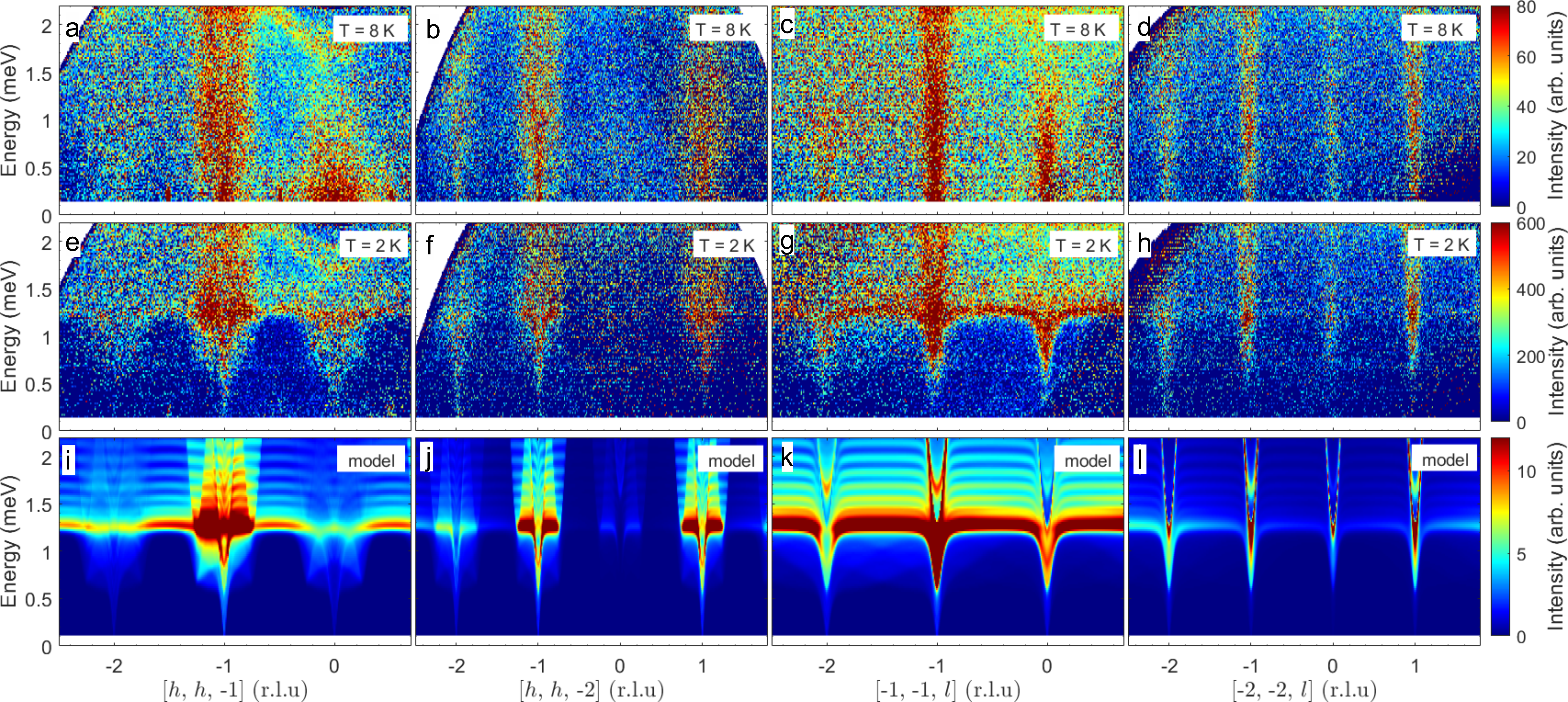}
\caption{Excitation spectra obtained on the single crystal of \scto along the $l$ direction using the time-of-flight spectrometer LET, with an incident energy of $E_{i}=2.76$~meV. The intensity includes $\pm0.25$~r.l.u in the vertical scattering plane ($\lbrack{h,-h,0}\rbrack$). \textbf{(a)}-\textbf{(b)} show the spectra along $(h,h,1)$ and $(h,h,2)$ directions at $8~K$, \textbf{(c)}-\textbf{(d)} show the spectra along $(1,1,l)$ and $(2,2,l)$ directions at $8~K$. Note:  The measurements at the two temperatures were performed with different neutron flux (and instrumental normalization) which resulted in approximately 8 times higher intensity for the spectrum at $2$~K for the same counting time. \textbf{(e)}-\textbf{(h)} show the modifications in the ordered state at $T=2$~K. Intensity in all the eight spectra is obtained by integrating $\pm0.25$~r.l.u along the in-plane orthogonal directions. \textbf{(i)}-\textbf{(l)} show the corresponding spin-wave spectrum calculated using linear-spin-wave theory from the $J_1-J_2$ interchain model described in the text.}
\label{Fig4}
\end{figure*}

To extract the weaker interactions that couple the chains leading to magnetic order in \scto, we now discuss the differences in the magnetic excitations in the ordered state at temperature $T=2$~K with respect to the higher temperature data at $8$~K. As the powder data discussed in Fig.~\ref{Fig2}(d)\&(f) suggests modification in the excitations below $2$~meV, the low-energy part of the spectrum of single crystal is obtained at the LET spectrometer with incident energy $E_i=2.76$~meV. Figure~\ref{Fig4}(a)-(h) presents excitation maps along the $[h,h,-1]$, $[h,h,-2]$, $[-1,-1,l]$ and $[-2,-2,l]$ directions above and below the magnetic transition. In addition to the magnetic signal, these spectra contain visibly spurious dispersive features at small wave-vector transfers and energy transfer  $E>1.7$~meV (see Fig.~\ref{Fig4}(a),(c),(e),(g)) due to the sample environment. No background subtraction has been performed on the spectra. Nonetheless, magnetic intensity is clearly distinguishable in these plots. Particularly, the spectra at $8$~K exhibits continuous streak-like features with intensity modulating between maximum and minimum at $h=2n+1$ and/or $l=2n+1$ for every $[h,h,l]$ position. This observation is consistent with the spinon continuum behaviour as discussed in Sec.~\ref{secB}. 

At $2$~K, additional features due to magnetic ordering are clearly visible in the spectra. While most of the spectral weight below $E<0.5$~meV is shifted to higher energies in all the spectra, presented in Fig.~\ref{Fig4}(e)-(h), a weak excitation intensity persists down to the zero energy transfer within the energy resolution. This is accompanied by a relatively sharper dispersion along the $[h,h,-1]$ and $[-1,-1,l]$ directions which flattens at $\sim 1.25$~meV (in Fig.~\ref{Fig4}(e),(g)). Additionally, the rod like continuum originating at $[-1,-1,-1]$ also develops a clear split above the flat band as shown in Fig.~\ref{Fig4}(g). Although the flat band is absent in the spectra along $[h,h,-2]$ and $[-2,-2,l]$ directions, a weak modification in the intensity is observed at this energy. These observations are consistent with the magnon-excitations expected from the spin-wave dispersion due to 3D magnetic ordering in \scto. However, it is notable that the excitations, although observable, are not very sharp. We believe this is due to the finite-width of the $S(\mathbf{Q},\omega)$ in the two perpendicular directions in addition to the finite energy resolution. To confirm this aspect and also to extract the magnetic interactions responsible for this low-energy modification, the dynamical structure factor is calculated using linear-spin-wave-theory on a simple Heisenberg model containing first three nearest neighbour interactions.

The general Hamiltonian to describe the magnetic properties of \scto can be written as:

\begin{equation}
\mathcal{H}=\mathcal{H}_{inter}+\mathcal{H}_{intra}
\label{eq2}
\end{equation}
where $\mathcal{H}_{intra}$ is according to the Eq.~\ref{eq1} and 
\begin{equation}
\mathcal{H}_{inter} =J_{1}\sum_{(i<j)}\mathbf{S}_{i}\cdot\mathbf{S}_{j}+J_{2}\sum_{(i<j)}\mathbf{S}_{i}\cdot\mathbf{S}_{j}
\label{eq3}
\end{equation}

Here, \jin$=J_3$ dominates the high-temperature magnetism giving rise to the multispinon-continuum and $J_{inter}$ (including $J_1$ and $J_2$) are responsible for the bandwidth and the dispersion of the spin-wave excitations in the low-temperature ordered state. To reproduce the low temperature, low energy spin-wave spectrum, it is insufficient to consider only the $\mathcal{H}_{inter}$. This is because, the boundary between the magnon and spinon contributions in a weakly coupled spin-1/2 antiferromagnetic chains is not well-defined giving rise to a crossover region as in the case of the prototypical compound KCuF$_3$~\cite{Lake2005}. Also, in spin wave theory the energy scale perpendicular to the chains is influenced by the strength of $J_{intra}$. Hence, it is essential to include the intra-chain term in calculating the spin-wave spectrum of the system to accurately represent the intensity and energy scale of the spinwaves. The lower boundary of the 2-spinon continuum in the chain direction follows the des Cloizeaux-Pearson dispersion which is renormalized upwards from the expected spin-wave dispersion and can be modeled by spin-wave theory using a fixed $J^{'}_{intra}=\frac{\pi}{2}$\jin$=6.63$~meV~\cite{Cloizeaux1962} in Eq.~\ref{eq2} in the simulation of spin-wave spectrum. 

\begin{figure}
\includegraphics[width=1 \columnwidth]{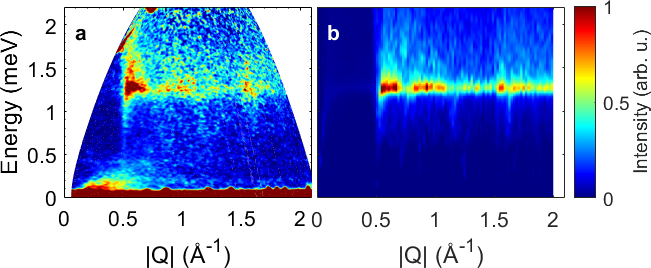}
\caption{\textbf{(a)} Excitation spectra of \scto obtained on the polycrystalline sample using the time-of-flight spectrometer LET, with an incident energy of $E_{i}=2.76$~meV. \textbf{(b)} The simulated spin-wave structure factor of a spin-1/2 Heisenberg antiferromagnetic chain considering $J_1=0.17$~meV, $J_2=-0.037$~meV and re-normalized $J_3^{'}=\frac{\pi}{2}J_3=6.63$~meV.}
\label{Fig6}
\end{figure}

The simulations are carried out using the SPINW software~\cite{Toth2015}. Here, the Cu$^{2+}$ form factor as well as the instrumental energy resolution are taken into consideration. The ground state of the Hamiltonian was fixed to the $120^{\circ}$ co-planar magnetic structure as described in the previous report~\cite{Chillal2020} and the four-dimensional neutron structure factor $S(\mathbf{Q},\omega)$ was simulated for several combinations of the interchain interactions. We find that the antiferromagnetic terms $J_1$, $J_2$ and $J_6$ as suggested by the DFT calculations in the previous report~\cite{Ahmed2015} are not compatible with the assumed magnetic structure and hence linear spin-wave theory completely fails. Instead, our simulations suggest a weak, finite antiferromagnetic exchange $J_1\sim 0.2$~meV is necessary to describe the bandwidth of the low-energy spin wave spectrum in addition to a much weaker ferromagnetic $J_2$. 
To obtain accurate values of these inter-chain couplings, the in-plane $S(\mathbf{Q},\omega)$ is scaled and fitted with the experimental data within SPINW resulting in the values: $J_1=0.17\pm0.025$~meV, $J_2=-0.037\pm0.012$~meV. Figure~\ref{Fig4}(i)-(l) show the simulated neutron structure factor for the spin wave excitations along the $[h,h,-1]$, $[h,h,-2]$, $[-1,-1,l]$ and $[-2,-2,l]$ directions considering the fitted values of antiferromagnetic $J_1$ and a much weaker ferromagnetic $J_{2}$. The structure factor also includes a finite width within the scattering plane in the perpendicular direction ($\pm0.25$~r.l.u) and also in the vertical scattering direction $[h,-h,0]$ ($\pm0.025$~r.l.u) just like the data. As we show in the appendix.~I, these contributions are necessary to predict the correct averaged-intensity of the experimentally observed spin-waves which also result in extra `out-of-plane' flat modes above energies of $1.25$~meV. However, the presence or absence of these weak modes cannot be conclusively verified in the data due to the high-background at these energy transfers compounded with the broadening of spin-waves with increasing energy. Nevertheless, the simulated results capture the main features of the experimental single crystal data as well as the powder average of the excitation spectrum as shown in Fig.~\ref{Fig6}.


Alternatively, the observed spin-wave dispersion can also be described by a purely ferromagnetic interchain coupling on $J_2=-0.11$~meV aided by an additional 6th neighbour interaction $J_6$ that couples the adjacent chains along the same cubic axes as shown in the fig.~\ref{Fig1}. While this interaction (either ferromagnetic or antiferromagnetic) may play a role in stabilizing the magnetic structure leading to the relative 90~$^{\circ}$ orientation of the spins in the parallel chains, it cannot be solely responsible for the observed three-dimensional magnetic ordering. Consequently, the magnitude of $J_6$ does not affect the energy scale of the spin-waves, however the structure is compatible when it is antiferromagnetic ($J_6=0.16$~meV) and comparable to $|J_2|$. The resulting spectra also produce an extremely similar pattern compared to the $J_1-J_2$ model with small differences in the intensities. Therefore, the $J_2-J_6$ model qualitatively agrees with the latest DFT Hamiltonian~\cite{Bag2021}, however, the magnitude $J_2$ found in our analysis is 4 times smaller. 

\section*{Discussion}

The fitted interactions of the Heisenberg Hamiltonian are found to be consistent with the 120$^{\circ}$ magnetic structure suggested by the neutron diffraction. As described in Fig.~\ref{Fig1}, this structure is made of 12 non-collinear Cu$^{2+}$ spins in the unit cell emphasizing the complexity of the system. Therefore, it is pleasantly surprising to find a simple model with linear spin-wave theory. The agreement with the data is achieved in both the energy scale and their intensities. In the experiment, the excitations appear broader especially at higher energies, this is expected as the spin-wave model is only really appropriate at lowest energies. Other modes not included in spin-wave theory add to the apparent broadening such as the longitudinal mode found in compounds with low ordered moment which is also expected at low energies~\cite{Essler1997,Lake2000}.

As expected from the large value of intra-chain interaction $J_3$, the magnetic excitation spectra of \scto is dominated by the multi-spinon continuum. On the other hand, two independent models are found for describing the low temperature spinwave excitations. In the $J_1-J_2$ model, the weak antiferromagnetic $J_1$ and ferromagnetic $J_2$ connecting the chains are responsible for stabilizing the commensurate magnetic structure at low temperatures and modifying the low-energy part of the excitations. We find that these values of the couplings are consistent with the sum of interactions predicted from the high temperature Curie-Weiss temperature in Ref.~\cite{Chillal2020b}. The ratio of the interchain coupling to the intra-chain coupling $\frac{J_{inter}}{J_{intra}}=\frac{J_1}{J_3}\sim\frac{1}{25}$ indicates that \scto is highly one-dimensional. 

The magnon excitations in the ordered state are gapless suggesting the predominant isotropic nature of the Cu$^{2+}$ spins. However, as discussed in Ref.~\cite{Chillal2020b}, application of external field induces anisotropic response in the magnetic phase diagram, particularly, parallel to the local ordered spin-direction $[h,h,0]$. This is also observed in the field-induced ($>2$~T) electric polarization~\cite{Koteswararao2015} suggesting the presence of weakly anisotropic terms. For example, DM interactions in \scto cannot be completely ignored due to the lack of inversion symmetry on all the relevant bonds responsible for $J_1$, $J_2$, $J_3$ and also $J_6$. Moreover, all the three components of DM are allowed for these bonds rendering the experimental extraction of the parameters complex. Nevertheless, if DM were to influence the magnetic structure of the system, its magnitude must be comparable to that of the experimental net interchain coupling $J_1$ ($-0.17$~meV) leaving the strongest $J_3$ bond as the likely candidate. This should lead to canting of the spins around their mean antiferromagnetic alignment along the chain. However, our diffraction results have shown that this is not the case. On the other hand, small values of DM on the weaker $J_1$ and/or $J_2$ are possible. We note that the addition of even a single component DM$_{x,y,z}$ on either of these bonds (off-diagonal elements of the symmetric $J_{1}$ or $J_{2}$) compatible with this magnetic structure will produce a gap proportional to its total magnitude $|$DM$|$ in the spinwave spectrum of \scto. This is valid for both $J_1-J_2$ as well as $J_2-J_6$ interchain models. Therefore, we estimate that $|$DM$|$ if present will be less than 0.01~meV, approximately 10\% of $J_1$ ($J_2$) in the simplest $J_1-J_2$ model ($J_2-J_6$ model).  However, input from improved DFT methods that include the spin-orbit coupling of the Copper spins will be extremely useful in verifying the antisymmetric elements of the Hamiltonian. Further theoretical approaches involving the quantum effects due to the strong one-dimensional character would also be helpful to resolve the role of $J_1$ more precisely. Recent DFT results exclude $J_1$ based on the length and bond angle aspects of the superexchange path in favour of the flatter sixth nearest neighbour $J_6$ for the same ground state energy involving antiferromagnetic $J_1$ and ferromagnetic $J_2$~\cite{Bag2021} [see appendix.II for a detailed comparison of DFT Hamiltonian with experimental data]. However, it should be noted that a similar scenario exists for \pcto where the magnitudes of $J_1$ and $J_2$ are predicted to be equally strong and antiferromagnetic.
 
\section*{Conclusion}
In summary, we have studied magnetic excitations of \scto in polycrystalline and single crystal samples and propose two Heisenberg magnetic Hamiltonians in the ground state. We find that chain interaction $J_3$ is the dominant magnetic exchange path in both models which gives rise to the diffuse spinon continuum. The simplest inter-chain coupling is led by an antiferromagnetic $J_1$ along with a weak ferromagnetic $J_2$ resulting in a three dimensional magnetic ordering of the Cu spins below T$_{N1}$ giving rise to  the sharper spin-wave features at low energies. Additionally a ferromagnetic $J_2$ and antiferromagnetic $J_6$ interchain coupling is also proposed.

\section*{Acknowledgements}
We thank Jean-S\'{e}bastien Caux for his calculation of the dynamical structure factor of the spin-1/2 Heisenberg antiferromagnetic chain. B.L acknowledges the support of the Deutsche Forschungsgemeinschaft (DFG) through the project B06 of the SFB-1143 (ID:247310070). The powder synthesis and crystal growth took place at the Core Lab for Quantum Materials, Helmholtz Zentrum Berlin f\"ur Materialien und Energie, Germany. 

\section*{Appendices}
\subsection*{I. Finite integration in perpendicular directions}\label{ap2}

\begin{figure}
\includegraphics[width=1. \columnwidth]{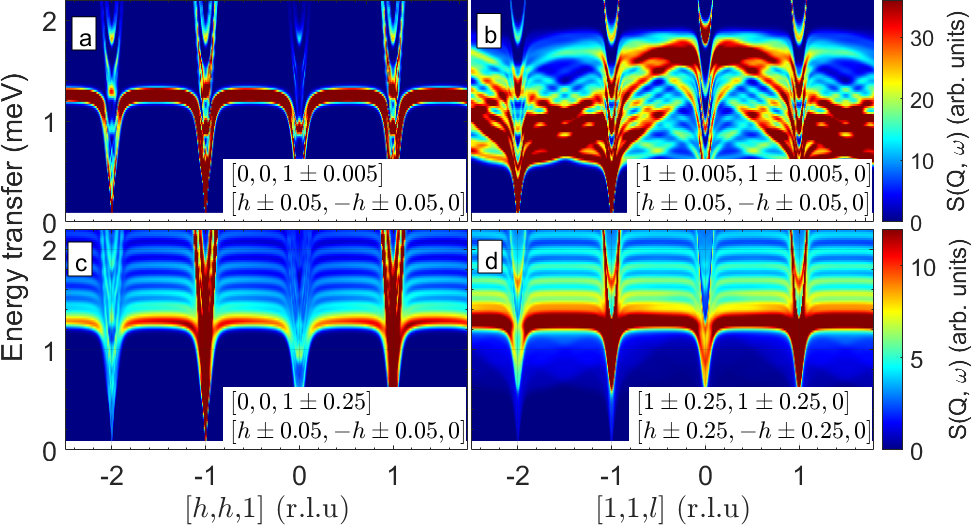}
\caption{Simulated spin-wave spectra of \scto without in-plane integration along \textbf{(a)} $[h,h,1]$,  \textbf{(b)} $[1,1,l]$. \textbf{(c)-(d)} show the spectra including finite contributions from orthogonal in-plane direction as well as vertical to the scattering plane.}
\label{Fig1a}
\end{figure}

The presence of three-dimensionally coupled three mutually perpendicular spin chains in \scto is readily evident in the intensities of the observed excitation spectra. Figure~\ref{Fig1a}(a)-(b) show the simulated spin-wave dispersions along $[h,h,1]$ and $[1,1,l]$ directions with minimal contribution from the in-plane orthogonal direction, namely $[0,0,l]$ and $[h,h,0]$ respectively. The plots also include a finite out-of-plane contribution along $[h,-h,0]$ direction. The spectrum perpendicular to the chain direction $l$ exhibits steep and sharp features throughout the Brillouin zone with several transverse modes at higher energies (see Fig.~\ref{Fig1a}(a)). On the other hand, the intensity along $[1,1,l]$ (see Fig.~\ref{Fig1a}(b)) reveals a complicated spectra indicative of the superposition of contributions from several modes. Therefore, it is expected that the excitation spectra will be strongly affected when even small contributions from other planes are included. In our scattering plane, this is revealed as additional flat bands above 1.25~meV, as shown in Figure~\ref{Fig1a}(c)-(d). We find that the number and the extension of bands towards higher energies is strongly affected by the integration width perpendicular to the slice.   

\subsection*{II. Comparison with density functional theory}\label{ap3}

\begin{figure}
\includegraphics[width=1. \columnwidth]{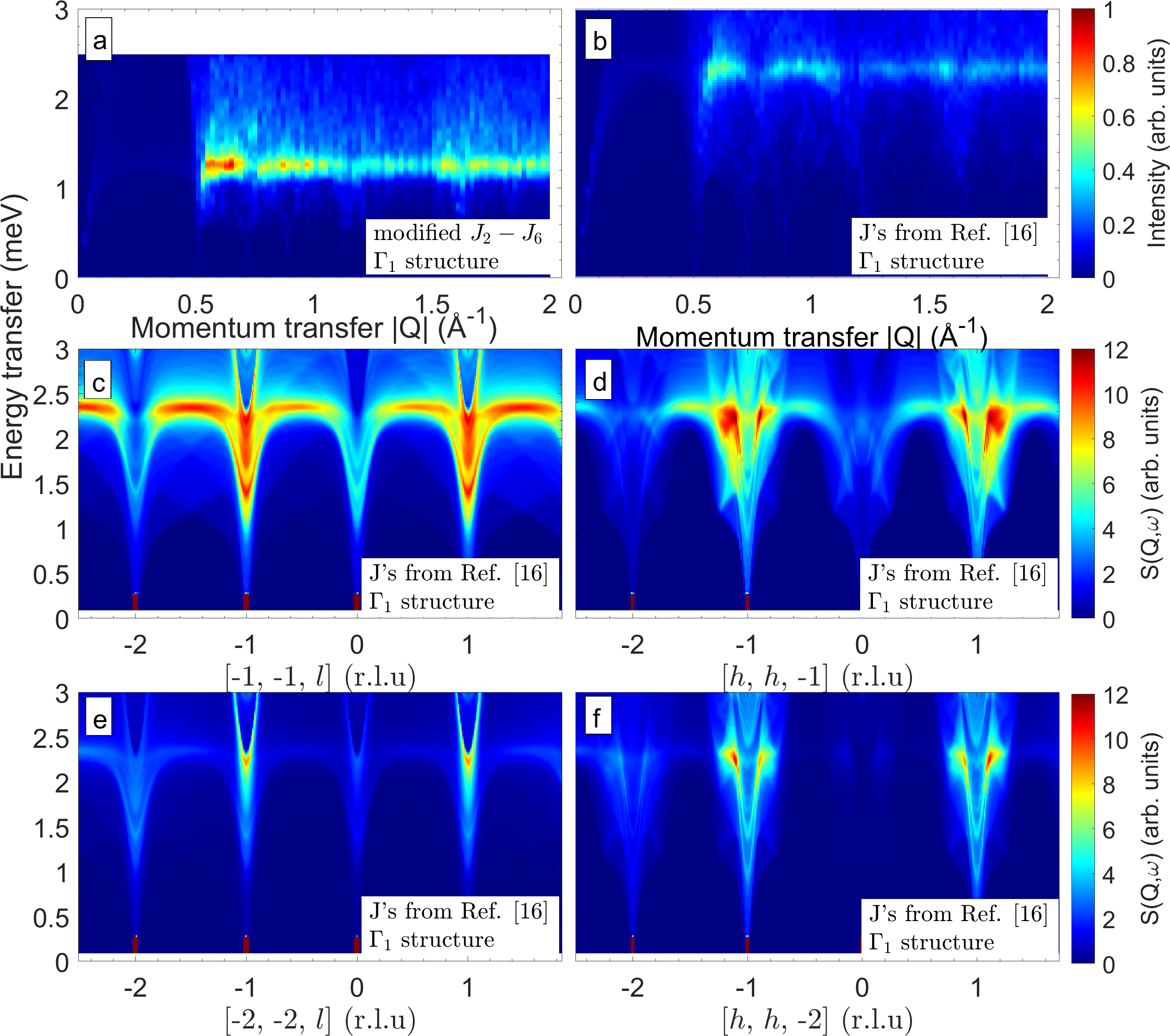}
\caption{Powder average spectra of \scto simulated from the $J_1-J_6$ Hamiltonian obtained in the \textbf{(a)} present work and, \textbf{(b)} from DFT calculations~\cite{Bag2021}. \textbf{(c)-(f)} Show the simulated momentum-resolved spin-wave spectra from DFT~\cite{Bag2021} along $[-1,-1,l]$, $[h,h,-1]$,$[-2,-2,l]$ and $[h,h,-2]$  directions for the same conditions as Fig.~\ref{Fig4}(i)-(l).}
\label{Fig2a}
\end{figure}

The relevant magnetic exchange interactions in \scto have been calculated using the density functional theory (DFT) by Ahmed et. al.~\cite{Ahmed2015} and bag et.al.~\cite{Bag2021}, both of which suggest $J_2$ as the leading interchain interaction followed by $J_6$ (with ratio ${|J_2|}/{|J_6|}\sim2$). When $J_2$ is antiferromagnetic ~\cite{Ahmed2015}, the Hamiltonian is incompatible with $\Gamma_1$ (120$^{\circ}$ $J_1$ triangular order) magnetic structure where linear spin-wave theory completely fails. However, the spin-wave dispersion and the corresponding structure factors can be calculated by using approximate diagonalization of the Hamiltonian in the ground state for a ferromagnetic $J_2$. The simulated excitation spectra from this Hamiltonian~\cite{Bag2021} including the experimental conditions for powder average as well as momentum-resolved single crystal sample are presented in Fig.~\ref{Fig2a}. When compared to the present work (see Fig.~\ref{Fig2a}(a)), the couplings from DFT clearly overestimate the energy scale of the spin-waves in powder sample of \scto (see Fig.~\ref{Fig2a}(b)). This is also evidenced in the simulated single crystal spectra plotted in Fig.~\ref{Fig2a}(c)-(f) along the four $[h,h,l]$ directions even though the overall features along multiple symmetry directions are in broad agreement with experimental data. Also, the weighted sum of the interactions from Ref.~\cite{Bag2021} result in a Curie-Weiss temperature of $\theta_{CW}\sim-41$~K, a considerably higher value compared to the experimental -27~K. Therefore, we find that the interchain couplings predicted by DFT are overestimated compared to the experimentally observed spin-wave excitation spectrum of \scto.

\bibliographystyle{apsrev4-2}
%
\end{document}